# Social tipping processes for sustainability: An analytical framework

**Authors**
*Ricarda Winkelmann[1,2]\*[†], Jonathan F. Donges[1,3]\*[†], E. Keith Smith[4,5]\*[†], Manjana Milkoreit[6][†], Christina Eder[4], Jobst Heitzig[7], Alexia Katsanidou[4,8], Marc Wiedermann[7], Nico Wunderling[1,2,9], Timothy M. Lenton[10]*

**Affiliations**
(1) FutureLab on Earth Resilience in the Anthropocene, Earth System Analysis, Potsdam Institute for Climate Impact Research, Member of the Leibniz Association, Telegrafenberg A31, 14473 Potsdam, Germany
(2) Institute of Physics and Astronomy, University of Potsdam, Potsdam, Germany
(3) Stockholm Resilience Centre, Stockholm University, Kräftriket 2B, 114 19 Stockholm, Sweden
(4) GESIS – Leibniz Institute for the Social Sciences, Unter Sachsenhausen 6-8, 50667 Cologne, Germany
(5) International Political Economy and Environmental Politics, ETH Zurich, Switzerland
(6) Department of Political Science, Purdue University, 100N University Street, West Lafayette, IN 47906, United States of America
(7) FutureLab on Game Theory and Networks of Interacting Agents, Complexity Science, Potsdam Institute for Climate Impact Research, Member of the Leibniz Association, Telegrafenberg A31, 14473 Potsdam, Germany
(8) Institute of Sociology and Social Psychology, University of Cologne, Cologne, Germany
(9) Department of Physics, Humboldt University of Berlin, Berlin, Germany
(10) Global Systems Institute, University of Exeter, Exeter, EX4 4QE, United Kingdom
*\* Corresponding authors: ricarda.winkelmann@pik-potsdam.de, jonathan.donges@pik-potsdam.de, keith.smith@gess.ethz.ch*
*† shared lead authorship*

**Abstract**
Societal transformations are necessary to address critical global challenges, such as mitigation of anthropogenic climate change and reaching UN sustainable development goals. Recently, social tipping processes have received increased attention, as they present a form of social change whereby a small change can shift a sensitive social system into a qualitatively different state due to strongly self-amplifying (mathematically positive) feedback mechanisms. Social tipping processes have been suggested as key drivers of sustainability transitions emerging in the fields of technological and energy systems, political mobilization, financial markets and sociocultural norms and behaviors.

Drawing from expert elicitation and comprehensive literature review, we develop a framework to identify and characterize social tipping processes critical to facilitating rapid social transformations. We find that social tipping processes are distinguishable from those of already more widely studied climate and ecological tipping dynamics. In particular, we identify human agency, social-institutional network structures, different spatial and temporal scales and increased complexity as key distinctive features underlying social tipping processes. Building on these characteristics, we propose a formal definition for social tipping processes and filtering criteria for those processes that could be decisive for future trajectories to global sustainability in the Anthropocene. We illustrate this definition with the European political system as an example of potential social tipping processes, highlighting the potential role of the FridaysForFuture movement.

Accordingly, this analytical framework for social tipping processes can be utilized to illuminate mechanisms for necessary transformative climate change mitigation policies and actions.

# MAIN TEXT

## 1. Introduction

There is growing concern that global climate change is reaching a point where parts of the Earth System are starting to pass damaging climate tipping points (*1*): In particular, part of the West Antarctic Ice Sheet (WAIS) appears to already be collapsing because of irreversible retreat of grounding lines (*2*, *3*) which in turn is expected to trigger loss of the rest of the WAIS (*4*). Other tipping points may be close: A recent systematic scan of Earth system model projections has detected a cluster of abrupt shifts between 1.5 and 2.0°C of global warming (*5*), including a collapse of Labrador Sea convection with far-reaching impacts on human societies. The abrupt degradation of tropical coral reefs is projected to be almost complete if warming reaches 2.0°C (*6*, *7*). The possibility of the global climate tipping to a 'hothouse Earth' state has even been posited (*8*).

Against this backdrop, there is a growing consensus that avoiding crossing undesired climate tipping points requires rapid transformational social change, which may be propelled (intentionally or unintentionally) by triggering social tipping processes (*9*, *10*) or "sensitive intervention points" (*11*, *12*). Examples for such proposed social tipping dynamics include divestment from fossil fuels in financial markets, political mobilization and social norm change, socio-technical innovation (*9–11*, *13*, *14*). Equally, if human societies do not act collectively and decisively, climate change could conceivably trigger undesirable social tipping processes, such as international migration bursts, food system collapse or political revolutions (*15*). Social tipping processes have received recent attention, as they encompass this sort of rapid, transformational system change (*9*, *10*, *13*, *15*).

Here we develop an analytical framework for social tipping processes. Drawing upon expert elicitation and a comprehensive literature review, we find that the mechanisms underlying social tipping processes are categorically different from other forms of tipping, as they uniquely have the capacity for agency, they operate on networked social structures, have different spatial and temporal scales, and a higher degree of complexity. Following these distinctions, we present a definitional framework for identifying social tipping processes for sustainability, where under critical conditions, a small perturbation can induce non-linear systemic change, driven by positive feedback mechanisms and cascading network effects. We adopt this framework to understand potential social tipping dynamics in the European political system, where the *FridaysForFuture* movement (*16*) pushes the system towards criticality, generating the conditions for shifting climate policy regimes into a qualitatively different state.

The proposed framework aims to establish a common terminology to avoid misconceptions, including the notions of agency, criticality as well as the manifestation and intervention time horizons in the context of social tipping. In this way, the framework can serve to connect literatures and science communities working on social tipping, social change, complex contagion dynamics and evidence from behavioral experiments (e.g. *14*, *17*).

## 2. Background

### 2.1. *Tipping points as social-ecological systems features*

We start by reviewing the characterization of tipping points across the natural and social sciences. Over the last 150 years, a suite of concepts and theories describing small changes with large systemic effects has been developed at the intersection of natural and social sciences. More recently, the concepts of tipping points and tipping elements have been broadly adopted by both natural and social scientists working within the field of climate change.

While the concept of 'tipping' originated in the natural sciences (*18*, *19*), social scientists made extensive use of the idea in the 20th century, often without using the terminology of tipping.

Famously, Schelling (*20*), following Grodzins (*21*), developed a theory of tipping processes to explain racial segregation in US neighbourhoods. Granovetter (*22*) modeled collective behavior as a tipping process that depends on passing individual thresholds for participation in riots or strikes. Kuran (*23*) described political revolution in terms of tipping dynamics, while Gould and Eldridge (*24*) distinguish phases of policy change and stability in terms of 'punctuated equilibrium'. Gladwell (*25*) popularised the concept of 'tipping points', exploring contagion effects ("fads and fashions"), sometimes triggered by specific events.

Several recent studies have examined tipping processes within contemporary social systems. Homer-Dixon (*26*) and Battison (*27*) explored the 2008 financial crisis as a tipping phenomenon. Nyborg (*14, 28*) discussed shifts in norms and attitudes, for example regarding smoking behaviors. Centola (*17*) associated tipping points with the "critical mass phenomenon", wherein 20–30% of a population becoming engaged in an activity can be sufficient to tip the whole society. Similarly, Rockström et al. (*29*) highlighted this so-called Pareto effect in the context of decarbonization transitions. Kopp et al. (*15*) distinguished different social tipping elements within the realm of policy, new technologies, migration and civil conflict that are sensitive to "climate-economic shocks". Here, a tipping element is a system or subsystem that may undergo a tipping process.

Since the mid 1990s, ecologists and social-ecological systems (SES) researchers have also developed an extensive body of research on tipping processes using the terminology of 'regime shifts' and 'critical transitions' (e.g. *30–32*). Recognizing the impacts of human development on various ecosystems, this body of work often models ecological regime shifts as a consequence of social drivers. Less attention, however, has been paid to sudden changes in social systems triggered by ecosystem changes.

There is a rich literature on the collapse of past civilizations (e.g. *33, 34*) and the potential role of tipping points in that (*35*). Recently, Cumming and Peterson (*36*) brought this together with work on ecological regime shifts, proposing a "unifying social-ecological framework" for understanding resilience and collapse. Further, Rocha et al. (*37*) noted that tipping dynamics can be produced by the interactions between climatic, ecological and social regime shifts.

The concept of climate tipping elements introduced by Lenton et al. (*1*) and Schellnhuber (*38*), has been increasingly adopted within Earth and climate sciences. Climate tipping elements are defined as at least sub-continental-scale components of the climate system that can undergo a qualitative change once a critical threshold in a control variable, e.g., global mean temperature, is crossed. Positive feedback mechanisms at the critical threshold drive the system's transition from a previously stable to a qualitatively different state (*1*). Other scholars, e.g., Levermann et al. (*39*), suggest a somewhat narrower definition of climate tipping elements by introducing additional characteristics, such as (limited) reversibility or abruptness. The tipping elements identified so far include biosphere components such as the Amazon rainforest (*40–42*) and coral reefs (*6, 7*), cryosphere components such as the ice-sheets on Greenland and Antarctica (*43*), and large-scale atmospheric or oceanic circulation systems including the Atlantic meridional overturning circulation (*44, 45*). Their tipping would have far-reaching impacts on the global climate, ecosystems and human societies (e.g. *8, 46*).

*2.2. Social Tipping*

In response to the concept of climate tipping points, social scientists are re-engaging with this concept yet again, creating an additional layer of tipping scholarship with an emphasis on the need for and possibility of deliberate tipping of social systems onto novel development pathways towards sustainability (e.g. *11, 47*). Scholars argue in particular that the rapid, non-linear change of social tipping dynamics might be necessary to speed up societies' responses to climate change, and to achieve the goals of the Paris Agreement. It is this element of acceleration, propelled by positive feedbacks, that makes the concept of tipping particularly interesting. For example, Otto and Donges et al. (*9*) reported expert elicitations identifying social tipping elements relevant for driving rapid decarbonization by 2050. Rapid-paced changes are a distinctive feature potentially differentiating tipping dynamics from many other forms of social change, including incremental (policy or institutional) changes, or more radical (socio-technical) transitions or societal transformations.

Over the last decade, the literature on deliberate transitions and transformations towards sustainability has expanded significantly, exploring the dynamics that lead to the reorganization of social, economic or political systems (e.g. *48*, *49*). In many ways, this literature and the emerging work on social tipping are interested in very similar phenomena: fundamental shifts in the organization of social or social-ecological systems - a movement from one stable state to another - including a change in power relations, resource flows, as well as actor identities, norms and other meanings (*48*). Transformations can be fast, but speed is generally not one of their defining characteristics.

This temporal feature of social tipping points - rapidity of change compared to the system's normal background rate of change - combined with the fact that tipping processes can be triggered by a relatively small disturbance of the system is motivating scholarship on leverage or 'sensitive intervention points', e.g. Farmer et al. (*12*), who identified such potentially high-impact intervention opportunities, e.g., financial disclosure, choosing investments in technology and political mobilization that may be key for triggering decarbonization transitions.

Based on a bibliometric and qualitative review of these various bodies of literature across the natural and social sciences, Milkoreit et al. (*10*) proposed the following general definition of (social) tipping: "the point or threshold at which small quantitative changes in the system trigger a non-linear change process that is driven by system-internal feedback mechanisms and inevitably leads to a qualitatively different state of the system, which is often irreversible." Milkoreit et al. (*10*) further noted there is a need to recognize and identify potential differences between climatic (or ecological) and social tipping processes to gain a deeper understanding of these phenomena.

3. **Methods and analytical structure**

Given this diverse and nascent field, there is a clear need for consensus as to what defines social tipping processes, as well as an understanding of how these processes are similar and diverge from dynamics in other non-social systems. Further, there are currently limited examples of social tipping elements in the context of sustainability transitions presented within the broader literature (*9*, *12*, *13*, *15*).

Here we explore the characterization of tipping processes within the natural and social sciences, examining how social and climate tipping processes are differently conceptualized. We draw upon a mixed qualitative methodological approach to illuminate these differences and key distinctions. Initially, core differences were identified and discussed via expert elicitation (*50*). A selected group of 25 experts from across the climate and social sciences were invited to take part in an expert elicitation workshop, that focused on identifying a common definition for social tipping processes, as well as the characterization of their dynamics. This workshop was convened in June 2018 in Cologne, Germany. The workshop participants were split into cross-disciplinary breakout groups, to independently identify the dynamics of social tipping processes. Then, each of these groups reported their findings to the broader plenary, for discussion, consolidation, reconciliation and clarification. The process was then repeated for further clarification within the breakout groups. Through this iterative inductive and deductive process, several unique themes and characteristics were identified from the broader set of codes, resulting in the key differences in and definition of social tipping processes presented below.

Drawing upon the differences identified in the expert elicitation workshop, we then review and synthesize the emerging field of social tipping processes, particularly in comparison to the related climate and ecological tipping dynamics. We then draw upon these unique characteristics to develop a common definition for social tipping processes, which we explore using the example of the *FridaysForFuture* student movement.

## 4. Results

### *4.1. Key differences between social and climate tipping processes*

Social and climate systems' tipping processes exhibit several broad, fundamental differences in their structure and underlying mechanisms: (i) agency is a main causal driver of social tipping processes, (ii) the quality of social networks and associated information exchange provides for specific social change mechanisms not available in non-human systems, (iii) climate and social tipping processes occur at different spatial and temporal scales, and (iv) social tipping dynamics exhibit significantly more complexity than climatic ones.

**Agency:** The most important characteristic differentiating social from climate tipping processes is the *presence of agency*. While a significant body of work (e.g. *51*), including Latour's actor-network theory (*52*), addresses different forms and effects of non-human or more-than-human agency, here, we focus on a more narrow understanding of agency that is based on consciousness and cognitive processes such as foresight, planning, normative-principled and strategic thinking, that allow human beings to purposefully affect their environment on multiple temporal and spatial scales. While humans have a generally poor track record of utilizing their agentic capacities especially with regard to shaping the future (e.g. *53–55*), they appear unique in their capacity to transcend current realities with their decisions.

Agency in this more narrow sense can be understood as the human capacity to exercise free will, to make decisions and consciously chart a path of action (individually or collectively) that shapes future life events and the environment (*56*). The notion of intentionality inherent in the idea of agency implies that human actors are not only able to adapt to changes in their environment, but also deliberately create such changes. Non-human life forms can also be engaged in deliberate changes of their environment (e.g., beavers building dams), but the cognitive quality of these actions differs from those of humans, which can be based on different forms of knowledge and meaning about the world, moral norms and principles, or ideas about desirable futures. Agency allows individuals and societies to be proactive rather than merely responsive in their relationships with other humans or the environment through planning, goal setting and strategic decision-making, which links decisions and behaviors in the present with consequences and realities in the (distant) future (*57*).

Governance scholars address this social-cognitive capacity for forethought and goal-pursuit in terms of anticipation (*58*) and imagination (*10*), which can be tied to a set of futuring methods (*59*, *60*). The ability to anticipate and imagine futures enables humans and their societies (*53*, *54*) – as opposed to animal communities or ecosystems – to transcend the present and shape the future according to our values and goals (*61*), possibly increasing the prospects for human survival in times of fast and significant environmental change (*56*, *62*). Although this ability has been underutilized in the past, especially in the context of responding to climate change (*63*), it is a crucial dimension of the human repertoire of tools to create change and to ensure its long-term well-being.

Agency interacts with many of the additional differentiating characteristics we identify below in important ways. For example, agency plays a role in the creation of social networks, institutions and meaning, i.e., the production of the structures of social systems. These network structures in turn enable and constrain agency (e.g. *64*, *65*).

Physical climate tipping elements, such as ice sheets or ocean circulations, lack that ability to intentionally act and adapt. However, the adaptive capacity of ecosystems can be interpreted as a form of non-human agency and learning mechanism (*66*), see also Supplementary Information S2. While scholarship on non-human agency, including that of animals, inanimate objects, landscape features or ecosystems (e.g. *67, 68*) might expand our understanding of agency, the cognitive abilities that characterize human agency, especially long-term and strategic thinking, do not exist in the non-human or inanimate worlds.

**Social networks:** Understanding the *nature of social networks* is crucial for studying social tipping. While both natural (including physical and ecological) and social systems can be structurally characterized as networks and studied using a network science approach (*69*), social systems differ from natural systems in the quality of the networks' nodes and interconnections and the processes and dynamics facilitated and impacted by these particular network characteristics. Social systems feature additional network levels of information transmission (cultural and symbolic) that are largely restricted to human societies compared to natural systems (*70*).

*Network qualities unique to social systems:*
Networks in social and natural systems share various commonalities such as the existence of fundamental nodes and links (*69*). In contrast to most natural systems, however, social networks have the capacity to intentionally generate new nodes, which include socially constructed entities such as organizations and movements (*71*). New nodes can be created through cultural, political or legal means, as can the rules for their interactions with other existing nodes. Social system nodes are unique in that they have richer cognitive realities, particularly agency and forethought. These nodes often have conflicting vested interests, which may be more short-sighted than future oriented.

Relationships in social networks can consist of shared meanings – especially norms, identities and other ideas – and a vast variety of cultural, economic and political relationships (e.g., employment, citizenship), all of which are not as pronounced or non-existent in less complex human societies and nature. Hence, social network links are more diverse than links in natural systems and enable different kinds of network processes. For example, links between nodes in social networks are not necessarily dependent on physical co-presence, due to technologically enabled connections or the presence of more abstract interrelations such as shared norms, values or interpersonal relationships.

*Network processes*:
Social network dynamics can be of a purely ideational nature (e.g., the subject of the study of opinion and belief dynamics), but also involve material changes (e.g., resource extraction, movement and transformation for economic purpose). Markets are unique social networks, involving both ideational and material network processes. In the Anthropocene, the intensity and speed of socially networked interaction has increased dramatically, largely due to new media, digitalization, more efficient means of transportation, lower travel costs, and overall increased mobility, which is likely to increase spreading rates, while at the same time affecting the stability of the network itself (*72–74*).

Generally, social tipping can either occur on a given network (e.g., through spreading dynamics changing the state of nodes (*75*) or change the network structure itself (see Figure 1). The structural network changes generated by social tipping processes include transitions from centralistic or hierarchical to more polycentric (neuromorphic) structures in urban systems, energy distribution and generation networks (*76, 77*). Structural changes can manifest on large and small-scale spatial networks across multiple social structure levels. In order to capture these network tipping processes, quantifiers from complex network theory such as modularity, degree distribution, centrality or clustering can be used (*69*).

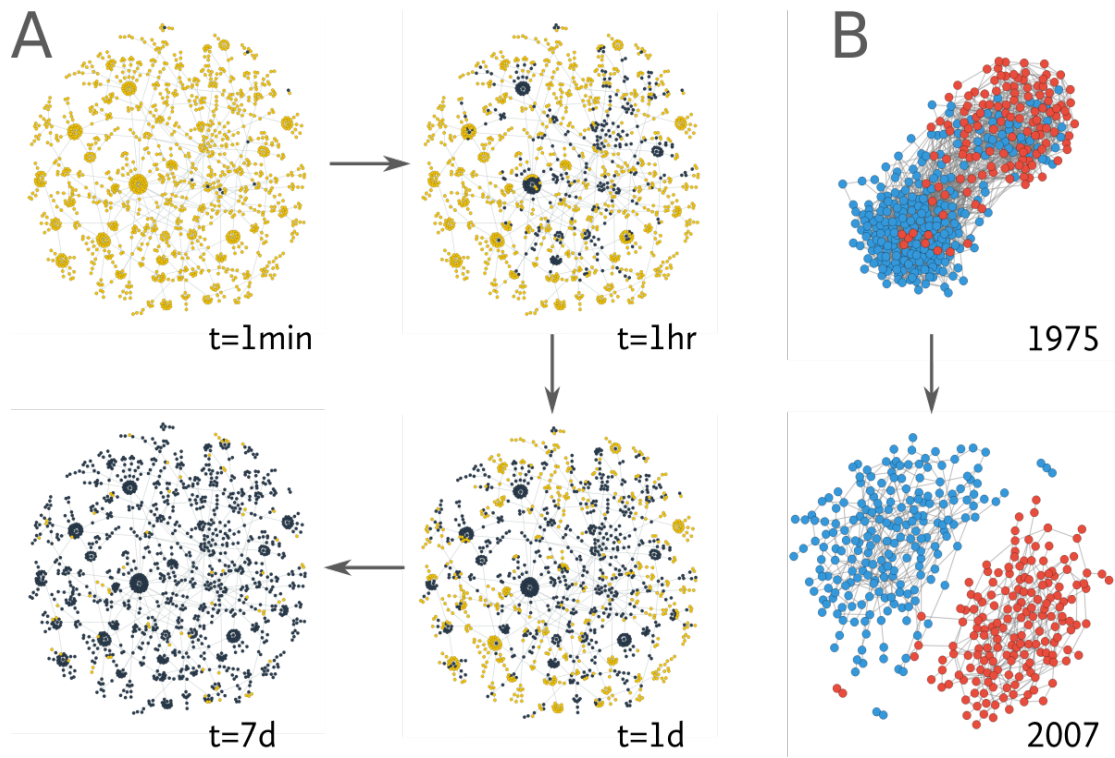

*Figure 1: Two types of social tipping in a complex network. (A) Social tipping can on the one hand be characterized by a contagion process where initially only a few nodes exhibit a certain property that then spreads through a large portion of the network. (B) On the other hand social tipping may also qualitatively alter the entire network structure from, e.g., a state with closely entangled nodes of different states to an almost or full disintegration of the network in smaller disjoint groups. The example in (A) shows the spread of an avatar among users in an online virtual world over the course of one week after it was first introduced by a small number of users (78). Nodes represent users and links represent the imitation of the avatar from one user to another. Yellow nodes denote users that have not picked up the avatar, while black nodes indicate those that did. (B) The upper network shows the members of the House of Representatives in the 94th United States Congress (January 3, 1975 to January 3, 1977). Node colors indicate different party membership and links between nodes are drawn if the corresponding members agree on 66% of all votes in the considered two-year period. The lower network shows the same for the 110th United States Congress (January 3, 2007, to January 3, 2009). The transition from a closely entangled to an almost fragmented topology indicates a polarisation between Democratic and Republican Party members over time (16).*

**Temporal and spatial scales:** Scales can differ greatly between social tipping and climate tipping processes and are more ephemeral for social tipping than for climate tipping.

Temporally, tipping in social systems manifests more commonly on the scale of *months to decades*, while for the climate tipping elements range from *years to millennia*. Human actors tend to focus on more short-term consequences or outcomes, as complex issues (such as climate change) with longer timeframes are often harder to assess (*79*). Within social systems, fund manager performance is evaluated quarterly, politicians often think in electoral cycles, business operates with annual or five-year forecasts, while individual practices and dispositions are constantly evaluated and reevaluated (*80–82*). In natural systems, however, it might take decades, centuries or even millennia for outcomes of change processes to become detectable (see Figure 2).

Both social and climate tipping elements can be ordered spatially (*1*, *39*, *83*), although social tipping elements cannot always be precisely located geographically. Social scientists and economists have long grouped systems and processes as existing on the macro-, meso- and micro-levels (or some variation thereof), whereby some social systems (e.g., financial markets, political systems, technologies) consist of interdependent subsystems existing on multiple spatial levels.

Social tipping processes can also display spatial-temporal *ephemerality*. While climate tipping elements have a known spatial extent and dimensionality (with often a comparable extent in latitude and longitude and a generally much smaller extent in altitude) and have persisted in their current stable state for thousands (if not millions) of years, social tipping processes do not have a spatial extent or effective dimensionality that is known ex-ante and they can emerge (move into a critical state) and disappear (move out of a critical state) over time.

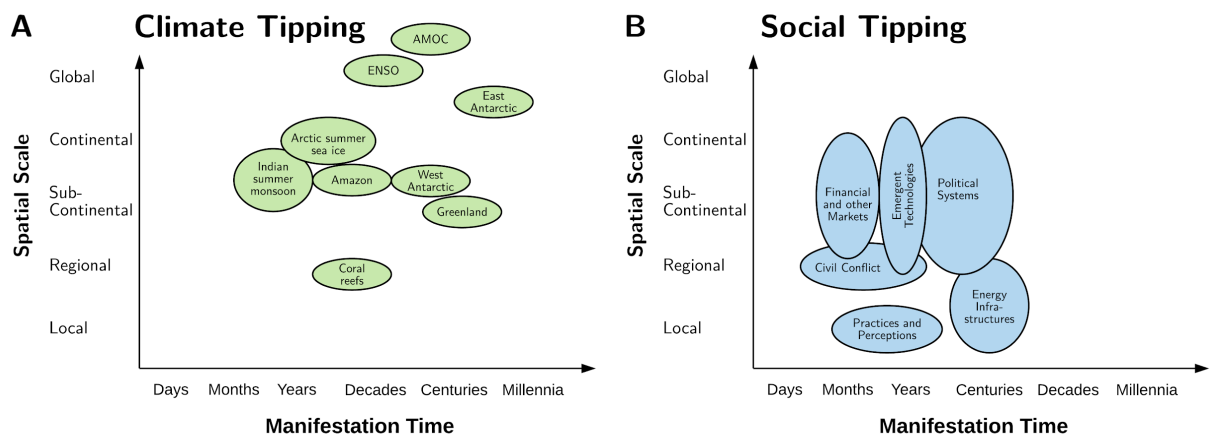

*Figure 2: Examples of spatial and temporal scales for climate and social tipping elements. Example climate tipping elements are broadly compiled from Lenton et al. (1), Levermann et al. (39), and Schellnhuber et al. (83). Social tipping elements are broadly compiled from Kopp et al. (15), Farmer et al. (12), Otto and Donges et al. (9), Hsiang (84), Tabara (11) and Lenton (13).*

**Complexity:** Social tipping processes occur in complex *adaptive* systems (*85–87*) as opposed to the complex but non-adaptive physical climate system. As such they can exhibit comparatively *greater complexity* in the (i) drivers, (ii) mechanisms and (iii) resulting pathways of social tipping processes, as well as the aforementioned ephemerality in their spatial-temporal manifestations, including a potentially fractal and varying dimensionality and a more complex interaction topology (*88*, *89*).

Social tipping processes can rarely be linked to a single common control parameter, such as is the case with global mean temperature in climate tipping dynamics. For most of the climate tipping elements like the ice sheets or the Atlantic meridional overturning circulation, the control variables such as local air temperature, precipitation or ocean heat transport, can often be translated or downscaled into changes in global mean temperature as one common driver (*1*, *38*). However, for social tipping processes, multiple, interrelated factors are often identified as forcing the critical transition. For example, shifts in social norms regarding smoking (*14*) can be linked to several, entwined factors, such as policies, taxation, advertising and communication, social feedbacks (e.g., via normative conformity), or individual preference changes. Centola et al. (*17*) show that tipping in social convention is possibly explained by a single parameter: the size of the committed minority). At larger scales, the collapse of complex civilizations has been linked to multiple interacting causes, and whilst disagreement abounds over the balance of causes in particular cases, there is general agreement that multiple factors were at play (*33*). This kind of causality – multiple interacting, distributed causes across varying scales – are a key characteristic of complex systems (*90*), contrasting starkly with conventional notions of causality involving bivariate relationships (one cause and one effect).

Further, due to their potential for agency and adaptive plasticity, social systems are open to a larger number of mechanisms that could cause a tipping process and various pathways of change that a tipping process could follow towards a greater number of potentially stable post-tipping states (*91*). Climate tipping processes are often modeled as bi- or multistable, where the directional outcomes of forcing are to some extent known or knowable, e.g., based on paleoclimatic data and process-based Earth system modelling. Given a specific forcing change, one can predict in what state the element will restabilize as well as the "net" effects of the tipping process on larger Earth systems. Based on this understanding, the tipping of climate system elements is generally perceived as undesirable and often as part of pushing the Earth system out of the "safe operating space for humanity" (*92*, *93*).

In contrast, for social systems, it is often unclear what a final stable state of the system will look like, or even whether the changes resulting from a tipping process will be normatively considered "positive" or "negative". As Clark and Harley (*94*) point out, the characteristics of complex-adaptive social systems, including the diversity of actors and elements and the different outcomes generated by local and global interactions, imply that the development pathways of these systems are less predictable. Further, a social tipping process can generate new and destroy existing actor types (e.g., identities, institutions) and their behaviors. Cross-scale dynamics and local differences are important to understand the emergent system structure and change dynamics, but predictive capacities, e.g., regarding the timing of a social tipping point or the boundaries between different stable states, do not yet exist (*94*). Hence, the term 'managing transitions' is less useful than the idea of navigating a transformation pathway.

The political nature of social change processes (*95*) – different actors within a social community pursuing different, sometimes opposing, interests and visions for a reorganization of a social system while bringing to bear different resources and strategies – further exacerbate this situation. Actors can deliberately generate new feedback dynamics that support or slow change, even after a tipping point has been passed, and they can actively work to adjust the direction of change.

### *4.2. Proposed definition of social tipping processes*

From the discussion above, it follows that a definition of social tipping process should take a micro-perspective and incorporate network effects and agency in addition to common tipping characteristics already explored in the review by Milkoreit et al. (*10*). It should also describe the timing aspects sufficiently well to understand possibilities for intervention, similar to what Lenton et al. (*1*) suggested for climate tipping elements. Hence we propose the following definition of the various terms relevant for studying social tipping processes (see Supplementary Material S1 for a more formal mathematical definition suggested for use in simulation modelling and data analysis that is consistent with what we put forward here):

> ***Definitions:*** *A 'social system' can be described as a network consisting of social agents (or subsystems) embedded within a social-ecological 'environment'. Such a social system is called a 'social tipping element' if under certain ('critical') conditions, small changes in the system or its environment can lead to a qualitative (macroscopic) change, typically via cascading network effects such as complex contagion and positive feedback mechanisms. Agency is involved in moving the system towards criticality, creating small disturbances and generating network effects. By this definition, near the critical condition the stability of the*

> *social tipping element is low. The resulting change process is called the '<u>tipping process</u>'. The time it takes for this change to manifest is the '<u>manifestation time</u>'.*[1]

If a tipping element *is* already in a critical condition, where the stability of its current state is low, there may be a time window during which an agential intervention might *prevent* an unwanted tipping process by moving the system into an uncritical condition (see also SI text S1). Alternatively, if a tipping element *is not* already in a critical condition, there may be a time window during which some intervention might move it into a critical condition in order to *bring about* a desired tipping process.

The small change triggering the tipping process could be either (i) a localized modification of the network structure (e.g., a change on the level of single nodes, small groups of nodes or links) or of the state of agents or subsystems, (ii) small changes of macroscopic parameters or properties, or (iii) small external perturbations or shocks. We deliberately do *not* require the trigger to be a *single* driving parameter. This is because we expect that a social tipping process could be triggered by a *combination* of causes rather than a single cause. Furthermore, a social tipping element may be tipped by several *different* combinations of causes. Consequently, for social tipping elements we cannot always expect at this point to identify a common aggregate indicator (such as global mean temperature in the case of climatic tipping elements) and a well-defined 'threshold' for this indicator at which the system will tip (see also the discussion on complexity above).

Note that social tipping as defined here is a unique form of social change, e.g., distinct from climate economic shocks (*15*) and more specific than socio-technical transitions (*96, 97*). Further, social tipping also denotes a shift to a qualitatively different state, and such, is different from standard business cycles or causes of seasonality. As such, social tipping presents a particular process of social change, where a system undergoes a transformation from one qualitatively different state to another, after being in a more critical state and affected by a potentially small triggering event.

### *4.3. Filtering criteria*

We propose several filtering criteria to focus on social tipping processes (i) that have the potential to be relevant to global sustainability in future Earth system tractories and (ii) where human interventions can occur within a pertinent *intervention time horizon* on the order of decades and will have consequences within a *political/ethical time horizon* on the order of hundreds of years.

(i) Relevance of social tipping for global sustainability

The social tipping process can impact a wide array of social systems, such as technological or energy systems, political mobilization, financial markets and sociocultural norms. We consider social tipping processes to be relevant here that have an impact on the biophysical Earth system or on macro-scale social systems. The qualitative change in a 'relevant' social tipping process significantly affects the future state of the Earth system in the Anthropocene directly or indirectly through interactions with other social tipping processes. Relevance can hence be defined in terms of impacts on biophysical Earth system properties such as global mean temperature, biosphere integrity or other planetary boundary dimensions. For example, tipping dynamics to a political system could result in policy regime changes, affecting substantial reductions in greenhouse gas emissions (*9, 12*). Furthermore, we consider social tipping processes that have relevant impacts on macro-social systems and can be triggered by changes in the same biophysical Earth systems, for example, mass migration due to climate impacts (*84, 98*).

---

[1] This is analogous to the 'transition time' in Lenton et al. (*1*) . We avoid the term 'tipping *point*' in this definition since some of the literature uses it to refer to a point in time while some of the literature uses it to refer to a certain state of the system or its environment.

(ii) Intervention and ethical time horizons

We are interested in potential social tipping processes in which humans have the agency to substantively intervene. For example, such interventions could be via technological or physical capacities of agential or structural actors. This therefore places emphasis on human intervention, such as decreasing the likelihood of extreme weather events via mitigation efforts, or triggering socio-technological changes towards decarbonization. We define intervention and ethical time horizons as follows:

*Intervention time horizon*
Human agency interferes with a social tipping element, such that decisions and actions taken between now and an 'intervention time horizon' could influence whether (or not) the system tips. We suggest to consider only social tipping processes with an intervention time on the order of 10 years (*9*), which arguably presents a practical limit of human forethought (*99*) and of future-oriented political agency. For example, international governance efforts for global sustainability challenges, such as the ozone regime or the Sustainable Development Goals, tend to work with similar time horizons. Similarly, social tipping processes for rapid decarbonization to meet the Paris climate agreement would have to be triggered within the next few years (*9*), with ambitious emissions reduction roadmaps aiming for peak greenhouse gas emissions in 2020 (*29, 100*). The intervention time horizon is analogous to the 'political time horizon' defined for climate tipping elements in Lenton et al. (*1*).

*Ethical time horizon*
The time to observe these relevant consequences should lie within an 'ethical time horizon'. This recognizes that consequences manifesting too far in the future are not relevant to the current discourse on how contemporary societies impact Earth systems. Such an ethical time horizon could consider only social tipping processes which can have relevant *consequences within the next centuries* at most, corresponding to an upper life expectancy of the next generations of children born.

## *4.4. Example of a potential social tipping process: European Climate Change Policy Dynamics Europe and FridaysForFuture*

Currently, international climate policies, including those of the European Union (EU) are insufficient to meet the +1.5°C or +2°C goals of the Paris Agreement (*101*). While European policy makers presume to lead global mitigation efforts and characterize their actions as ambitious (*102, 103*), actual policy measures and proposals have been lagging behind this aspiration (*104*). EU countries emit about a tenth of the world's emissions, and a policy change towards more rapid decarbonization would not only have significant direct impacts on the climate system, but likely have indirect effects on the policies of other major emitters. But what kinds of sociopolitical processes can lead to these necessary changes? Could such changes result from social tipping dynamics?

Public opinion is a crucial factor in policy formation, where the public can be understood as a "thermostat" signalling what is politically feasible (*105, 106*). Shifts in public opinion can punctuate previously stable and 'sticky' institutions, leading to policy change (*107*). Increased activism and public concern regarding climate change can generate new coalitions, or shift the priorities of existing ones (*108, 109*). Here we examine the European political system as an example of and how social tipping processes could be triggered as a result of large-scale public activism and social movements.

The European political system is composed of networks of agents (i.e., activists, decision-makers and organizations) with a range of social and political ties and is structured in nested and overlapping subsystems (i.e., national group, transnational political coalitions). Viewed through the lens of social tipping, European political dynamics present a 'social system', embedded within the broader international political and climate change governance community 'environment'. Driven by the *FridaysForFuture* movement (*16*) (among other things), a groundswell of bottom-up support for more proactive climate policies has recently developed among European citizens, resulting in routine mass demonstrations and historical wins for Green parties in the 2019 European Parliamentary Elections, as well as in federal elections in Austria, Belgium and Switzerland. The European political system could be moving towards a critical 'state', creating the conditions for a tipping process towards radical policy change, bringing European climate policy in line with the Paris Agreement. Accordingly, the European political system could constitute a potential 'social tipping element', where as it nears critical conditions, a small change to the system or its broader environment could lead to large-scale macroscopic changes, affected by cascading network dynamics and positive feedback mechanisms. Such transformations could involve establishing more aggressive mitigation strategies that connect goals (such as remaining below +2°C, 50% emissions reductions by 2030, zero carbon emissions by 2050) with measures and pathways that have a reasonable chance to achieve them (i.e., investment in negative emission technologies, increased carbon taxation policies etc.).

The *FridaysForFuture* movement has been pushing the European political system towards criticality, where it becomes more likely that the system will be propelled into a qualitatively different state. The movement was set off and inspired by a single Swedish high school student choosing to protest on the steps of the Riksdag for meaningful climate action. Greta Thunberg's protest quickly spread through the European social-political networks until more than a million students have been participating in weekly protests. This growing bottom-up pressure on the European climate policy-makers (*16*, *110*) has created an opening for significant policy change.

The European political system consists of embedded subsystems at multiple scales. At the national scale, for example, the German socio-political system responded strongly to the activities of the *FridaysForFuture* movement. Polling throughout 2019 in Germany suggested that the environment was the most important public policy challenge, ahead of other issues, such as the migration and financial crises. Drawing upon survey data collected monthly by the Politbarometer, 40–60% of Germans responded that the environment was an important problem in the Fall of 2019, a rapid increase from roughly 5% in the Fall of 2018 (Figure 3, Panels A and B). Since 2000, rarely more than 10% of Germans have viewed the environment as an important problem – a time period which includes the emergence of other large environmental movements in Germany, such as protests against nuclear energy in response to Fukushima. The specific upward shift in Germans viewing the environment as an important problem appears to coincide with the large-scale protests organized by *FridaysForFuture* in March, May and September of 2019.

Similarly, several national Western European Green Parties received historically strong electoral support in the May 2019 European Parliamentary Elections (such as in Belgium, Germany, Finland, France and Luxembourg). This increased support is also reflected in polling data in Germany, where the Green Party has been effectively equal with the conservative party as the preferred political party of German voters in the latter half of 2019 (Figure 3, Panels C and D). Subsequently, Germany introduced its first ever federal climate change laws, mandating that the country meet its 2030 goals (a ~55% reduction in GHG emissions) and establishing pathways to carbon neutrality by 2050. Currently, only a limited set of countries have enacted national climate change laws, and Germany is one of the largest and most diverse economies to propose such actions. This presents the possibility for policy diffusion and transfer to other states (*111*), particularly considering the influential role Germany plays within the European Union. Climate policy entrepreneurs could build upon momentum to further capitalize on windows of opportunity, pushing climate change proposals prominently into national and supra-national governmental agendas before the ephemeral moment passes (*112*).

The 2020 COVID-19 pandemic has placed new priorities on the policy agenda, also reflected in issue salience of climate change (see also Fig. S1 in Supplementary Materials). As political and behavioral responses to COVID-19 have led already to a significant temporary reduction in greenhouse gas emissions (*113*), this shock could be further leveraged to reinforce climate action – future economic recovery packages should set European economies on a pathway towards carbon-neutrality, rather than return to the old normal (*114*, *115*). Drawing from this social tipping framework, the European political system may remain near a critical state. It remains unclear whether the COVID-19 shock has supplanted climate change, or whether both remain on the political agenda. For example, discussions of a "Green New Deal" remain at the core of COVID-19 economic recovery plans within the European Union.

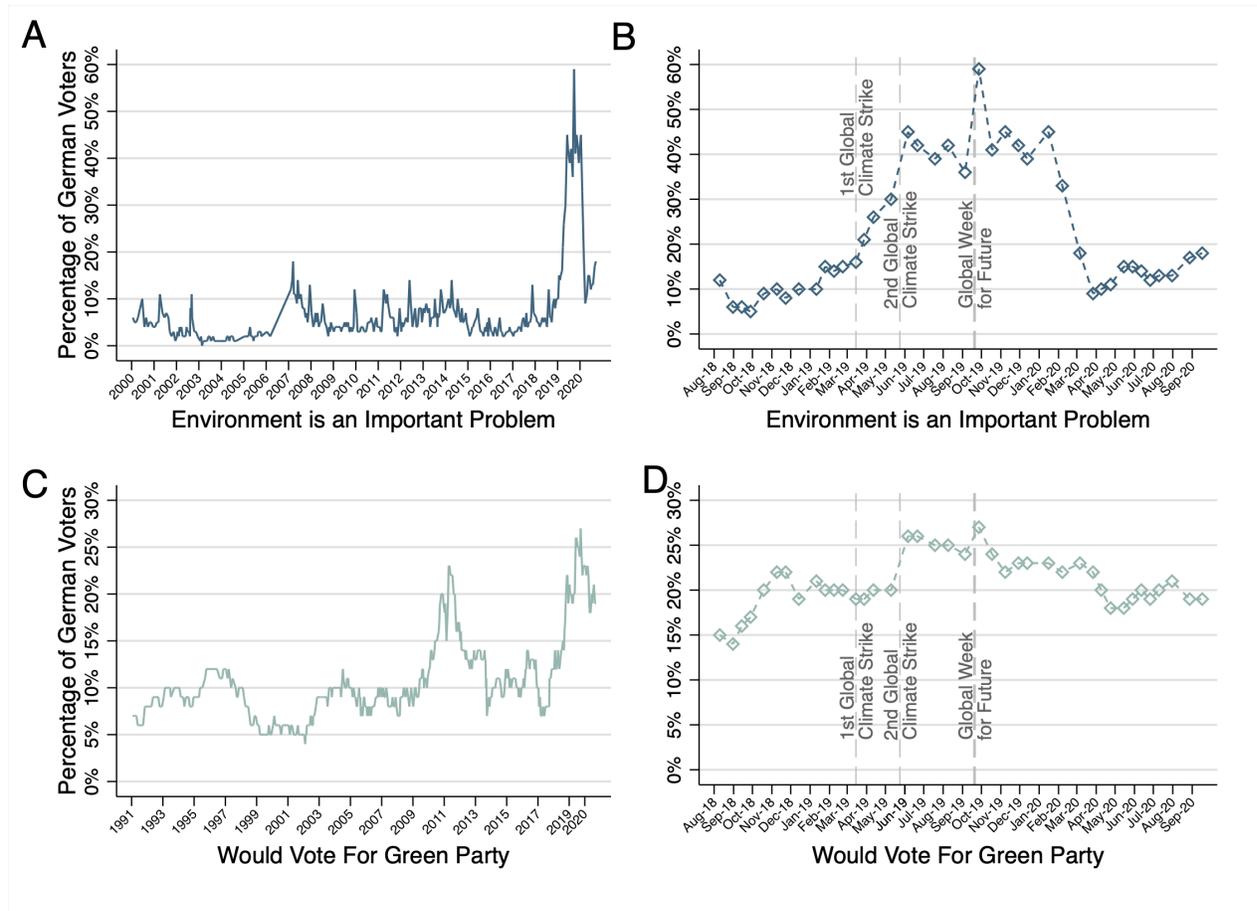

*Figure 3: Environment as an issue and willingness to vote for the Green Party in Germany. Percentages of potential German voters that list the environment as an important issue for the country and willingness to vote for the Green Party (Bündnis 90/Die Grünen) if the election were to be held "today". Panels (A) and (C) present monthly survey data from 2000 to September 2020 Panels (B) and (D) display monthly surveys from August 2018 – September 2020, showing the change since the beginning of Greta Thunberg's protest actions. Dotted grey vertical lines display days of global strikes organized by FridaysForFuture in March, May and September 2019. Data is collected by Forschungsgruppe Wahlen: Politbarometer .*

*Implications for criticality*

The sociopolitical dynamics have likely moved the Germany political subsystem further towards criticality, but it remains largely unknown whether this will result in tipping towards a qualitatively different state, in Germany or in the broader European political system. Rather, these judgements can *likely only be made in hindsight*, observing whether the system remained stable, moved towards criticality or experienced tipping dynamics. Such an analysis in line with the proposed framework requires specific process tracing, identifying the key moments, actors, networks, mechanisms affecting criticality, the triggering event (threshold), and the positive feedback dynamics propelling the system towards qualitative changes. Much attention is often paid to the specific triggering event, but it is rarely one single actor or action which accounts for the entirety of the tipping process.

Rather a full account needs to be made of all of the previous and related processes that have further placed the system towards criticality, allowing for such changes to become more likely. Accordingly, for a tipping process to occur at the scale of the entire European political system, moving it into a state of decarbonization that is aligned with the Paris Agreement, a series of additional social movements and protests, or other shifts within the system or the environment, may be required.

While we identify the role of *FridaysForFuture* in creating critical conditions, or potentially triggering the social transformations required for global sustainability, recent literature has identified further tipping candidates which could have generally "positive" effects on global sustainability. For example, divestment and reinvestment present candidates for rapid decarbonization and processes to achieve climate targets (*9*, *12*). In this case, intervention times range from years to decades, depending on the social structure level (*9*). Previous studies note that the adoption of technologies and behaviors such as rapid uptake of autonomously driven electric vehicles (if socially licensed), rapid change in dietary preferences reducing meat consumption and associated land-use and climate impacts can follow an epidemic-type model of diffusing across social networks (*13*, *15*).

Alternatively, social tipping processes can lead to states of criticality with less desirable outcomes: Recently it has been shown that climate change has contributed to the emergence of infections carried by mosquitoes, like dengue fever or Zika, which could be accelerated further by increased mobility, e.g., through denser air traffic networks (*75*). The thermal minimum for transmission of the Zika virus could in fact give rise to a threshold behaviour (*116*). Changes to the local environment may enact "push" factors, resulting in large scale migrations (*117*, *118*). Further, increased global mean temperature has been suggested to increase the likelihood of civil conflicts (*84*).

These social tipping processes are of great interest to policy makers, as it is desirable to potentially trigger or facilitate "positive" tipping (*11*, *13*), while at the same time, mitigating the effects of potential "negative" outcomes.

5. Discussion

Social tipping processes have been recently recognized as potentially key pathways for generating the necessary shifts for sustainability. Drawing upon this emerging field, this paper develops a framework for characterizing social tipping processes. We find that mechanisms underlying social tipping processes are more likely to exhibit the unique characteristics of agency, social-institutional and cultural network structures, they occur across different spatial and temporal scales to climate tipping, and the nature of tipping can be more complex. Social tipping processes thus present qualitatively different characteristics to those shared by climate tipping processes.

Accordingly, this paper develops a common framework for the unique characteristics of social tipping processes. We identify social tipping as a process, resultant of a complex system of drivers, resulting in shifting a system into a more (or less) critical state. It can thus serve to structure and inform future data analysis and process-based modelling exercises (*118*, *119*).

Even so, while there is an emerging focus on social tipping dynamics (*9–13*), there remains great difficulty in pinpointing tipping events and generalizing the emerging dynamics. Drawing from natural tipping dynamics, previous work on social tipping has often focused on identifying specific trigger events or critical thresholds in macroscopic system variables in analogy to identifying for instance critical temperature thresholds in the context of climate tipping (*10*). In natural systems the underlying dynamics are more deterministic and often can be directly observed, allowing for the identification of specific thresholds and events. While social systems comprise a much more open and complex system, one that is constantly adapting and where dynamics are often incredibly complex, interrelated and cannot be directly observed. Accordingly, one could observe the same event across ten similar social systems, and could potentially observe ten unique outcomes. As such, anticipating a specific trigger, making causal inferences, or having generalizability in expected effects are all greatly limited within social systems. Further, social tipping points are sometimes also understood as a point in time, rather than a point in a complex parameter space. Such an approach makes it difficult to identify social tipping processes, as they often do not contain easily observable macroscopic thresholds nor temporal markers for change.

Rather, a complex adaptive systems viewpoint is required, understanding the multitude of interrelated processes and social structures driving change, and not focusing on a single trigger or threshold. Accordingly, our framework proposed here focuses on identifying the processes and mechanisms of such change, and not a single triggering event, where the interplay of micro-level changes embedded within adaptive structural conditions can affect systemic changes.

The notion of a critical state is central within our framework. Changing conditions to the system's environment can cause it to enter more (or less) critical states, such that a single, or multiplicative action, can effect a systemic change. It is these changing conditions, and specifically the processes and dynamics underlying them, that are of analytical importance. Drawing upon the analogy of a tipping coal wagon (*15*), it is not the single, specific piece of coal that caused the wagon to tip, but rather the processes by which the wagon was filled with enough coal that any single piece (placed at a number of different locales) could cause such tipping. Accordingly, the specific triggering event of a social tipping process could be somewhat random or arbitrary, as the conditions are critical enough such that any event with enough magnitude could have triggered these dynamics.

It is therefore key to focus on the processes and mechanisms underlying the nature of such critical states which allow some trigger event to cause contagion dynamics. From social network models, we can deduce which kind of structural features make a system less resilient and thus more prone to social tipping (*119*). One example is polarization, where social network models and social media-based data analyses have shown that in polarized states with nearly disconnected network communities which in themselves are highly connected, contagion processes are more likely to occur (*120–122*). Behavioral experiments and corresponding conceptual modelling approaches suggest that minority groups can initiate social change dynamics in the emergence of new social conventions (*17*, *119*). Furthermore, a rich social science literature has noted an array of factors (i.e. political institutions, technological or behavioral adaptation, environmental, normative and attitudinal) effective in shifting the social conditions surrounding climate change (*14*). A better understanding of critical states as demanded by our framework may help to identify early warning signals that could possibly indicate that a social-ecological system is close to a critical state in specific situations (*30*, *123*).

Social tipping processes present a specific type of social change – characteristized by non-linear shifting states driving by positive feedbacks – which is similar to, but conceptually distinct from, other forms of social change. Similar to how we explore the differences between natural and social tipping processes, further research should engage with social tipping in comparison to other forms of social change (such as historical institutionalist perspectives, social movements, policy feedbacks, complex systems). One of the greatest challenges lies in dealing with multiple, entangled drivers of tipping processes on different scales – temporal, spatial or social structural levels – and different levels of agency and heterogeneous agents and subsystems. In order to further understand the dynamics arising from these various levels of agency, it is crucial to identify examples from different subfields (economics, political science, demographics). A key current limitation in applying our framework is finding and operationalizing empirical data describing actual spreading processes on networks across these different levels, particularly compared to macro-economic data and public opinion polls (*124*), even though first steps in this direction are being made (*125*, *126*). Particularly data on the social structures and networks is notoriously difficult to access. While there have been advances in developing modeling frameworks (*119*, *127*) to simulate social tipping dynamics, linking these theoretical modelling to empirical data and behavioral experiments requires more attention. Even if predictive modeling (i.e., the kind of deterministic, time-forward modeling we know from Earth System Models for instance) of such social dynamics in the sense of inferring time trajectories is very difficult or even conceptually unfeasible, such process-based modelling of social tipping dynamics can be very crucial to understand the nature of critical states also in real-world social situations. Lastly, we focus here specifically on social tipping processes relevant for mitigating climate change, or sustainability more broadly, fitting within the previous literature. But, such a framework for social tipping dynamics is generalizable to other areas of study and social phenomena (such as the 2020 rapid social movements and public opinion dynamics surrounding racial inequality in the United States).

While we explore one example of social tipping in detail, further inquiry is required to test the distinctiveness of social tipping processes, as well as the utility of the proposed definition to other social tipping processes. Systematizing the types of social tipping processes, and exemplary case studies, would help to further illustrate these forms of change. Research is also warranted into establishing typical timescales of social tipping; understanding how network structures affect social tipping dynamics; identifying typical network structures of systems entering critical states; discerning the temporal aspects of how effects travel through different social network structures; and gaining a better understanding of the origin of spreading processes. Data acquisition, analysis and process-based modelling could all play a role in this research agenda. A wealth of social media data is available to study potential social tipping processes. However, this kind of data has mostly yet to be adopted within the context of Earth System analysis and tipping dynamics.

Social tipping processes could be decisive for the future of the Earth System in the Anthropocene: some rapid shifts in social systems are, in fact, necessary to meet the targets of the Paris Agreement and the Sustainable Development Goals (*8*). While we focus here on processes relevant for future trajectories of the Earth system, we suggest that further analysis could use or adapt our definition to characterize other types of general social tipping processes (i.e. revolutions or rapid transformations). We also recognize that tipping processes within ecosystems present an interesting intermediary case between social and physical climate tipping as they typically incorporate characteristics from both realms. They are also crucial in determining future trajectories of the Earth system (see preliminary discussion in the SI). Understanding, identifying and potentially instigating some social tipping processes is highly relevant for the future of the Anthropocene, particularly with regard to the potential role in triggering rapid transformative change needed for effective Earth system stewardship (*9, 11–13*).

**Acknowledgments**

**General**: We are very grateful to William C. Clark, Anne-Sophie Crépin, Niklas Harring, Matthew Ives, J. Doyne Farmer, Wolfgang Lucht, and John Schellnhuber, for providing helpful insights and comments. We thank the participants of two DominoES workshops on social tipping dynamics held at GESIS Leibniz Institute for Social Science, Cologne, in summer 2018 and 2019 for foundational and framing discussions.

**Funding:** This work has been supported by the Leibniz Association (project DominoES), the Stordalen Foundation via the Planetary Boundary Research Network (PB.net), the Earth League's EarthDoc programme, IRTG 1740 funded by DFG and FAPESP, and the European Research Council (ERC advanced grant project ERA: Earth resilience in the Anthropocene, ERC-2016-ADG-743080). T.M.L.'s contribution was supported by the Leverhulme Trust (RPG-2018-046).

**Author Contributions:** Drawing upon the concepts developed in the expert elicitation workshop, R.W., J.F.D., E.K.S., M.M. and T.M.L structured the conceptualization into the resultant framework and wrote the paper with the support of all co-authors. All co-authors contributed to the discussion of the manuscript. M.W. analyzed data and created Fig. 1. R.W. and E.K.S. created Fig. 2. E.K.S. analyzed data and created Fig. 3. J.H. derived the mathematical definition of social tipping processes (Sect. S1).

**Competing Interests**: The authors declare no competing interests.

## Supplementary Materials

### S1: A mathematical definition of social tipping processes

In this section, we give a more formal version of the definition of 'social tipping process' given in the main text, as a reference for mathematically inclined readers.

After defining what we mean by a social system and its environment, we first classify their possible states into critical, unmanageable, uncritical, and tippable conditions, and then finally define the notions of prevention time and triggering time.

By a _social system_, $\Sigma$, we mean a set of agents together with a network-like social structure, that interacts in some form with the rest of the world, called the _environment_, $E$, of the system, such that, if no "perturbation" or deliberate "influence" by some decision-maker occurs, $\Sigma$ and $E$ together can only follow certain "quasi-inertial" (or "default") trajectories restricted by the agency of the system's agents. Let $x_{(t)}$ and $y_{(t)}$ denote the _states_ that $\Sigma$ and $E$ are actually in at time $t$.

A _critical condition_ for the system is a pair of possible system and environment states, $(x^*, y^*)$, such that there exists another possible pair of states, $(x', y')$, with the following properties:

1. The state pair $(x', y')$ is no further away in state space from $(x^*, y^*)$ than a certain "small" distance, $\epsilon$, that represents the possible magnitude of "local" perturbations in $\Sigma$ (affecting only few agents or network links directly) or small changes in $E$ that are considered sufficiently "likely" to care about, with respect to some suitable distance function $d$. In other words, $d((x', y'), (x^*, y^*)) < \epsilon$.
2. If $\Sigma$ and $E$ were in state $(x', y')$ at any time $t'$, there is a quasi-inertial trajectory that would move $\Sigma$ at some later time $t'' > t'$ into some state $x''$ that is "qualitatively" different from $x^*$. This move represents a "global" (i.e., affecting a very large fraction of the agents) and "significant" change in the system (but not necessarily in its environment).

If such a change actually happens, the time point $t'$ (not the state!) at which it starts may be called the _tipping point_ or less ambiguously the _triggering time point_, and the system behavior within the time interval from $t'$ to $t''$ is called the corresponding _tipping process_. An _uncritical_ condition for $\Sigma$ and $E$ then is any pair of states that is not critical.

A critical condition is _unmanageable_ for an actor that may influence $\Sigma$ or $E$ in some way if there exists a possible pair of states, $(x', y')$, with $d((x', y'), (x^*, y^*)) < \epsilon$ and the following property:

- Assume that $\Sigma$ and $E$ were in state $(x', y')$ at any time $t'$ and afterwards the state of $\Sigma$ and $E$ would follow any trajectory $(x(t), y(t))_{t \geq t'}$ that the actor can force it to follow. Then the resulting trajectory would still move $\Sigma$ at some time $t'' > t'$ into some state $x''$ (which will usually depend on the influence exerted) that is qualitatively different from $x^*$.

Similarly, an uncritical condition, $(x°, y°)$, is *tippable* by a decision maker if there is a possible trajectory $(x(t), y(t))_{t \geq t'}$, starting in $(x°, y°)$ at some time $t'$, that the decision maker can force $\Sigma$ and $E$ to follow, and this trajectory would move $\Sigma$ into some state $x''$ at some time $t'' > t'$ that is qualitatively different from $x°$ (a tippable uncritical state roughly corresponds to what others call a 'sensitive intervention point' ).

At any time at which the system is not in an unmanageable critical state, the *prevention time* is the time interval it takes before some quasi-inertial trajectory has moved it into an unmanageable critical state. In other words, at time zero it is the largest time interval $T$ so that, when no intervention takes place until time $T$, for all $t > 0$ with $t < T$, the system would not be in an unmanageable critical state at time $t$.

Similarly, at any time at which the system is in a tippable uncritical state, the *triggering time* is the time interval it takes before some quasi-inertial trajectory has moved it into an uncritical state that is no longer tippable. In other words, at time zero it is the largest time interval $T$ so that, when no intervention takes place until time $T$, for all $t > 0$ with $t < T$, the system would not be in a tippable uncritical state at time $t$.

We only consider social tipping processes for which the prevention or triggering time is smaller than some *intervention time horizon*.

## S2 Ecosystem tipping as intermediary case

Ecosystem tipping processes share properties of physical climate tipping dynamics in atmosphere, ocean and cryosphere in that they can often be described by a common driver, as well as that of deliberative social tipping elements in that they have adaptive capacity, and can therefore be regarded as intermediate. But, as previously noted, human agential capacity is far greater than those of other species.

Similarly to human social systems, ecosystems are comprised of interacting living organisms, they can be viewed as networks with components that can adapt (e.g., food webs). This is different from physical tipping elements such as the cryosphere elements (e.g., melting of permafrost) which do not typically exhibit the same networked structures. Within the nominally 'climate' tipping elements are some major biomes – notably boreal forests, the Amazon rainforest, and coral reefs – that are composed of living organisms and exhibit ecological network structures. Indeed changing interactions between the living elements of these systems may be key to tipping dynamics – for example epidemic bark beetle infestation of boreal forests triggered by climate warming allowing the beetles to complete two life cycles rather than one within a season (*128*). Thus these biotic tipping elements lie towards smaller scale ecosystems in the continuum, and tend to be more closely related to social systems in spatial and temporal scales compared to the typically much larger and more slowly changing physical climate tipping elements.

These differences give rise to a proposed ordering of tipping elements, ranging from (1) the physical climate tipping elements via (2) ecosystem tipping elements to (3) social tipping elements (Table S1).

*Table S1: Proposed ordering of tipping processes ranging from physical climate tipping processes via ecosystem tipping processes to social tipping processes.*

| Properties | Physical climate tipping processes | Ecological tipping processes | Social tipping processes |
|---|---|---|---|
| Degree of agency | *Low/Absent* | *Intermediate* | *High* |
| Network structure | *Uncommon* | *Common* | *Common* |
| Temporal-spatial scales | *Slower and larger* | *Faster and smaller* | *Faster and smaller* |
| Degree of complexity | *Lower* | *Intermediate* | *High* |

*Figure S1:*

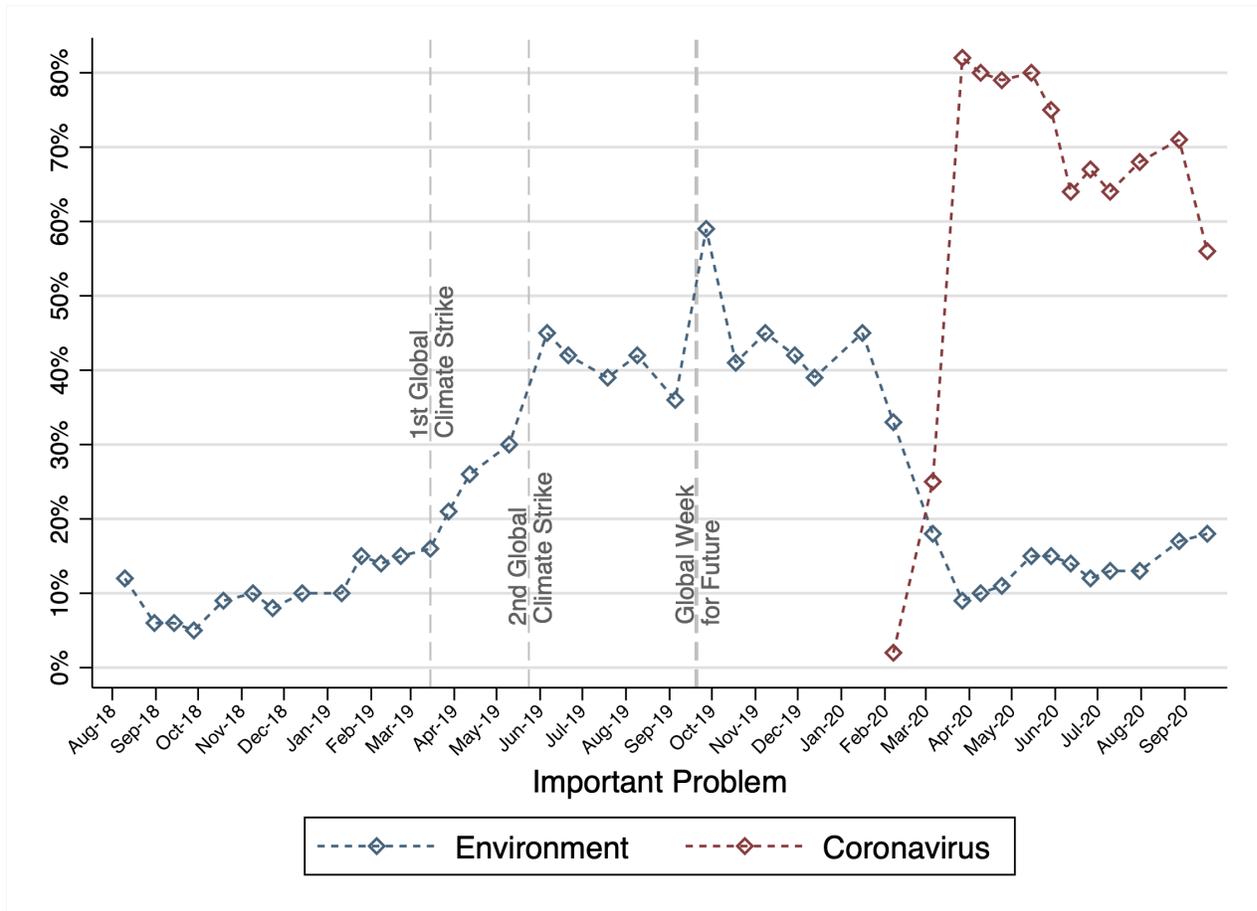

*Figure S1: Environment and Corona as an important issue in Germany.* Percentages of potential German voters that list the environment and the Coronavirus as an important issue for the country

*from August 2018 – September 2020, showing the change since the beginning of Greta Thunberg's protest actions. Dotted grey vertical lines display days of global strikes organized by FridaysForFuture in March, May and September 2019. Data is collected by Forschungsgruppe Wahlen: Politbarometer .*